\documentclass{article}[12pt]
\usepackage{amsmath}
\usepackage{amssymb}
\usepackage{amsthm}
\usepackage{amsfonts}
\usepackage{color}
\usepackage[pdftex]{graphicx}

\def\abstract#1{\vskip 7mm
        \begin{center}{\large Abstract}\par \smallskip
                \begin{minipage}[c]{12cm}
                        \small #1
                \end{minipage}
        \end{center}
}

\def\title#1{\begin{center}{\Large\bf #1}\end{center}}
\def\author#1{\vskip 5mm \begin{center}{#1}\end{center}}
\def\address#1{\begin{center}{\it #1}\end{center}}
\newcommand{\ssmatrix}[4]%
{\begin{pmatrix} #1 & #2 \\ #3 & #4 \end{pmatrix}}
\makeatletter
\@ifundefined{lesssim}{}{}
\@ifundefined{gtrsim}{}{}
\def\vereq#1#2{\lower3pt\vbox{\baselineskip1.5pt \lineskip1.5pt
\ialign{$\m@th#1\hfill##\hfil$\crcr#2\crcr\sim\crcr}}}
\makeatother

    {\mbox{}\hfill\mbox{\rm\normalsize$\Box$}\\%
    }
\usepackage{color}
\input{colordvi.tex}
\begin{document}

\title{Dynamical deformation of 2+1 dimensional double torus universe}

\author{Masaru Siino}

\address{Department of Physics, Tokyo Institute of Technology, Tokyo 152-8551, Japan}

\abstract{In (2+1)-dimensional pure gravity with cosmological constant, the dynamics of double torus universe with pinching parameter is investigated.
Each mode of affine stretching deformation is illustrated in the context of horizontal foliation along the holomorphic quadratic differential.
The formulation of the Einstein Hilbert action for the parameters of the affine stretching is developed. Then the dynamics along one holomorphic quadratic differential will be solved concretely.}

\section{Introduction}
Thurstone's geometrization conjecture, which insists that a compact three-dimensional manifold is decomposed into the building block admitting one geometrical structure among eight types of homogeneous geometry, was considered on the way to resolve the famous Poincar\'{e} conjecture.
Since Gregori Perelman proved the geometrization conjecture using Ricci flow and surgery, all the candidates of compact universe have been topologically classified for investigation in general relativity.
The first task was investigation of the so called homogeneous anisotropic geometry, the most of them are well studied in the context of Bianchi type geometry\cite{sol}, while almost studied their global degrees of freedom \cite{Koike} is not complete in several cases, where its origin is in two dimensional surface.

In two dimensions, the geometrization corresponds to the uniformization theorem\cite{unif} of Riemann surfaces, which classifies compact surfaces by its curvature into a sphere (positive), a torus (zero) and higher genus hyperbolic surfaces (negative). Since the geometry of the compact Riemann surface containing the global degrees of freedom as Teichm\"{u}ller deformation, its gravitational dynamics of the Riemann surface was studied by \cite{HNM}\cite{SF} especially in torus, but still hyperbolic higher genus surface is not completely understood, while its abstract formulation have been suggested\cite{HN2}\cite{SK}\cite{KP}.
Then since the three dimensional geometrization include product space of fibration on hyperbolic Riemann surfaces, this is also a part of the dynamics of compact three dimensional manifold which is not completely understood about the dynamics of Teichm\"{u}ller deformation of the hyperbolic Riemann surface.

In studying the dynamics of hyperbolic Riemann surface, there would be two different policies to characterize its configuration space.
Fenchel-Nielsen coordinate, which is globally defined based on the pants decomposition for `hyperbolic' structure, gives simple description of the phase space\cite{KP} but not suitable to analyze local geometry along the Einstein theory.
On the other hand, holomorphic quadratic differential, which is based on `flat' structure of the Riemann surface, is analytically defined from local deformation of flat metric and convenient to investigate the geometrical feature of the deformation for the Riemann surface.
Then, in order to study the geometry of compact hyperbolic space, that is the higher genus Riemann surface, we will analyze its complex structure and its complex differential as holomorphic quadratic differential.

Indeed, in (2+1)-dimensional gravity \cite{HNM}, the transvers traceless mode in ADM-formalism is identified to the holomorphic quadratic differential in the context of the complex structure.
Though the concrete calculation of the holomorphic quadratic differential is not well investigated, in the formulation of the pinching parameter there are several useful advantages.
The pinching parameter describes how two Riemann surfaces are connected by sewing on a pinching narrow bridge. In the expansion by small pinching parameter, Fay and Yamada\cite{YA} using Weierstrass elliptic function provided a formulation for a unique bilinear two form and holomorphic 1-form.
According to it, we have discussed\cite{MS1} the affine stretching of double torus to realize the degrees of freedom of Teichm\"{u}ller deformation (quasi-conformal mapping) and then homogeneous standard hyperbolic metric for a Riemann surface corresponding to the affine stretching along a holomorphic quadratic differential.
The homogeneous standard metric is regarded directly as geometry of the universe incorporating the degrees of freedom of global deformation, that is the dynamical variable of Teichm\"{u}ller parameter.

In the present article, we will analyze the dynamical deformation of the double torus.
The deformation mode along each holomorphic quadratic differential are illustrated, and concrete gravitational action and equation of motion are formulated.
As an example, we will see how the Teichm\"{u}ller variables and a scale factor arise in a system limited on a single mode of the holomorphic quadratic differential.

In the next section, introducing the result of our previous work, we give a concrete calculation of holomorphic quadratic differential and illustrate it in the context of horizontal foliation.
In the third section, we briefly remind the ADM-formulation and identify its transverse-traceless mode holomorphic quadratic differential (which is written HQD in brevity) in complex structure. Then we discuss its ADM-dynamics in the context of Weil-Petersson metric of configuration space using homogeneous standard metric for a double torus.
The final section is devoted to summary and discussion.

\section{affine stretching and horizontal foliation}

In our previous work\cite{MS1}, according to the references \cite{YA} we calculated the HQD of the double torus in the context of pinching parameter.
To understand the relation between the HQD and the geometry of the double torus universe, we will concern the affine stretching of the complex structure.
Incorporating the Teichm\"{u}ller deformation into the spatial metric tensor, we consider a spatial coordinate transformation with a parameter which will depend on time.
That is $z\mapsto z+k\bar{z},\ \ \ (0<k<1)$ in a local coordinate at a point and extending it by 
$z=h(\xi)$. Then the pull back of the metric tensor $dzd\bar{z}$ is $|dz+kd\bar{z}|^2=|h'd\xi+k\bar{h}'d\bar{\xi}|^2$.
In terms of HQD $\varphi=h'^2d\xi^2$, it reduces to 
\begin{align}
dl^2=|\varphi(\xi)|\left|d\xi+k\frac{\overline{\varphi(\xi)}}{|\varphi(\xi)|}d\bar{\xi}\right|^2,
\end{align}
where $\varphi=\varphi(\xi)d\xi^2$ behaves as $\varphi'(\xi')=\varphi(\xi)(d\xi^2/d\xi'^2)$ in coordinate transformation.

From the Riemann-Roch theorem, $\varphi$ spans complex $3g-3$ dimensional ($g$ is the genus) space $\langle\varphi^I\rangle_{\mathbb C}=A_2(R)$. Since $A_2(R)_1=\{\varphi\in A_2(R)|\ ||\varphi||_1<1\},\ \ \ (||\varphi||_1= 2\int_R|\varphi(z)|dxdy)$ is homeomorphic to the Teichm\"{u}ller space\cite{IT}, we regard the complex coefficients $Q^I$ of linear combination $\varphi=\sum_I Q^I \varphi_I$ as the Teichm\"{u}ller parameter for linear basis $\varphi^I$ of $A_2(R)$.

Geometrical meaning of that will be illustrated by   horizontal foliation defined by the HQD $\varphi_I$.
Given one HQD, it provides a holomorphic 1-form defining a local complex coordinate
\begin{align}
dq_I=\sqrt{\varphi_I}   .
\end{align}
In extending the complex local coordinate $q_I$ it generally does not gives global extension of $q_I$, by homotopically non-trivial integral, but only we can extend real axis of $q_I$ by line integration of 1-form $dq_I$, since $\varphi^I$ is holomorphic on the Riemann surface, as long as $\varphi_I$ does not vanish.

\begin{figure}[htpb]
\centering
\includegraphics[width=13cm]{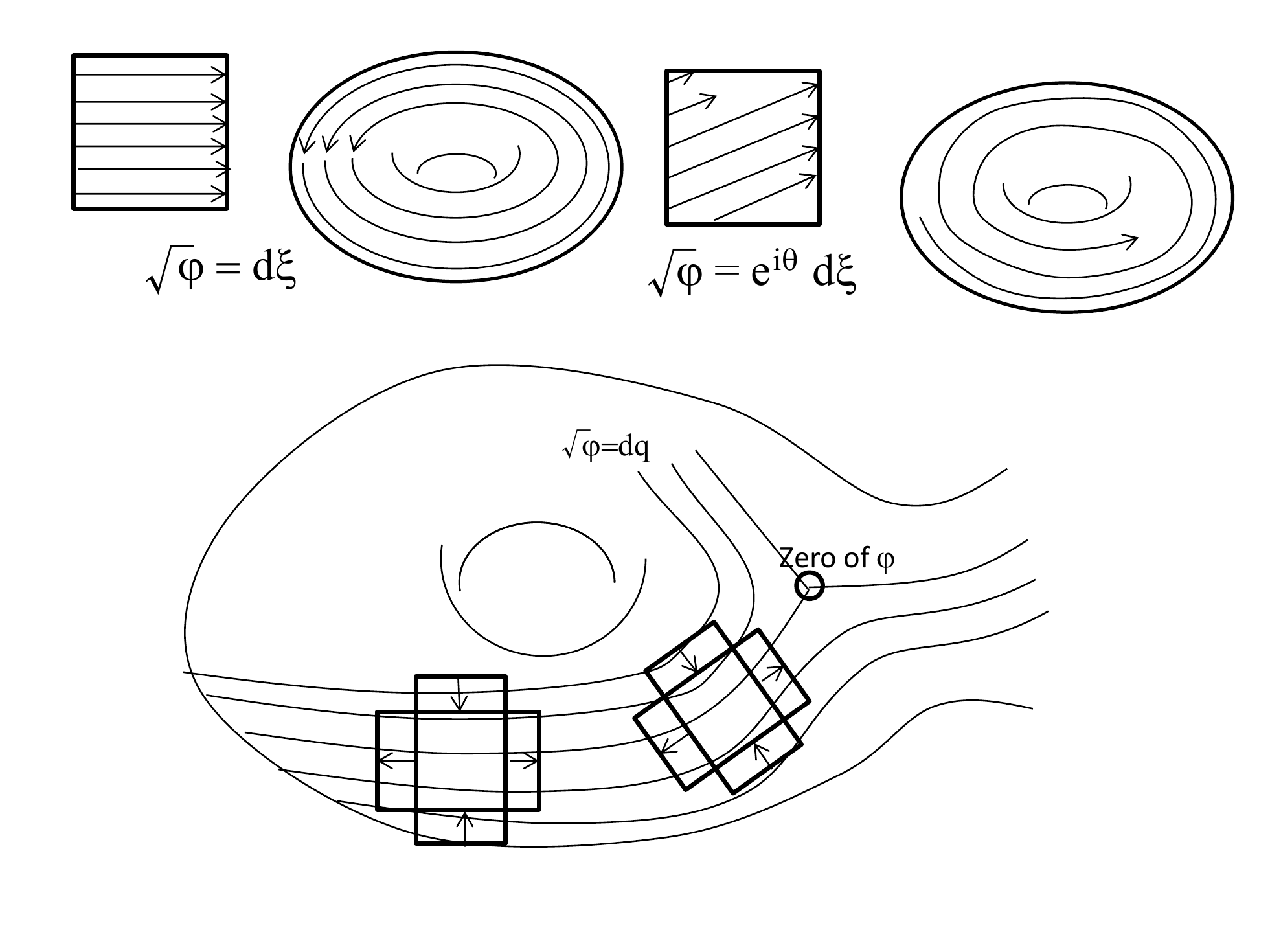}
\caption{Two typical cases of the horizontal foliation of the torus are shown in upper tori. In the left torus, the HQD is parallel to the global coordinate $\xi$. In the right case, that is not so and its horizontal foliation is non-trivial. In lower figure, a HQD has a zero of order one and its horizontal foliation has a trifurcation there.  The affine stretching is realized along the horizontal foliation.}
\label{fig:hortor}
\end{figure}

In the case of torus, $\sqrt{\varphi_I}$ is a complex constant $c\ dz$ and typical two cases are illustrated in Fig\ref{fig:hortor}.
When the boundary of fundamental region of the torus is parallel to the real axis, i.e. $c=1$, the integrated curves of the real axis is simply closed and can become a global coordinate, while the integral curve is repetitively moving around in the fundamental region for the case of $c=e^{i\theta}$ where the real axis is not parallel to the boundary of the fundamental region.
Anyway we have a foliation of the surface by such integral curves of the real axis, which is named as horizontal foliation.
Of course, for a higher genus Riemann surface, it is known that HQD $\varphi$ has $4g-4$ zeros.
The horizontal foliation is known to bifurcate at the zero, according to the order of the zero as illustrated in Figure \ref{fig:hortor}.

From the viewpoint of this horizontal foliation, the affine stretching is carried out in this local coordinate
$q_I$. Indeed, in the local coordinate $\varphi=d q_I^2$ is constant.
Then the coordinate transformation of the affine stretching is simply a linear transformation
$q_I'=(1+k)x-i(1-k)y  \ \ \ (q_I=x+i y)$ ($(1+k)$-stretching in horizontal direction and $(1-k)$-contraction in vertical direction), similar to the case of torus.
Thus by determining the horizontal foliation of $\varphi^I$, we will realize the Teichm\"{u}ller deformation (of quasi-conformal mapping) of the Riemann surface.

\subsection{horizontal foliation for a double torus of two tori sewn together}
Now we consider a horizontal foliation for a double torus of two tori sewn together by pinching parameter $\epsilon$.

It may be a consensus that it is very difficult to give a concrete expression of complex differential forms for  general Riemann surfaces with higher genus $g>1$, that is  normalized differential of the second kind  $\omega(u,v)$, holomorphic 1-form $\nu_i$, holomorphic quadratic differential  $\varphi_{ij}$.
On the other hand, for a genus $g$ Riemann surface with a narrow bridge structure, one can approximately compose the complex differential forms $\omega(u,v), \nu_i, \varphi_{ij}$ from the complex differential forms on two Riemann surfaces with lower genus, from which the genus $g$ Riemann surface is obtained by sewing.
Therefore, we only consider the case where a Riemann surface can be thought of as one composed of two Riemann surfaces connected by a narrow bridge. 
A general method for calculating $\omega(u,v)$ for any two sewn Riemann surfaces has been given by Fay and Yamada~\cite{YA}.

\begin{figure}[htpb]
\centering
\includegraphics[width=13cm]{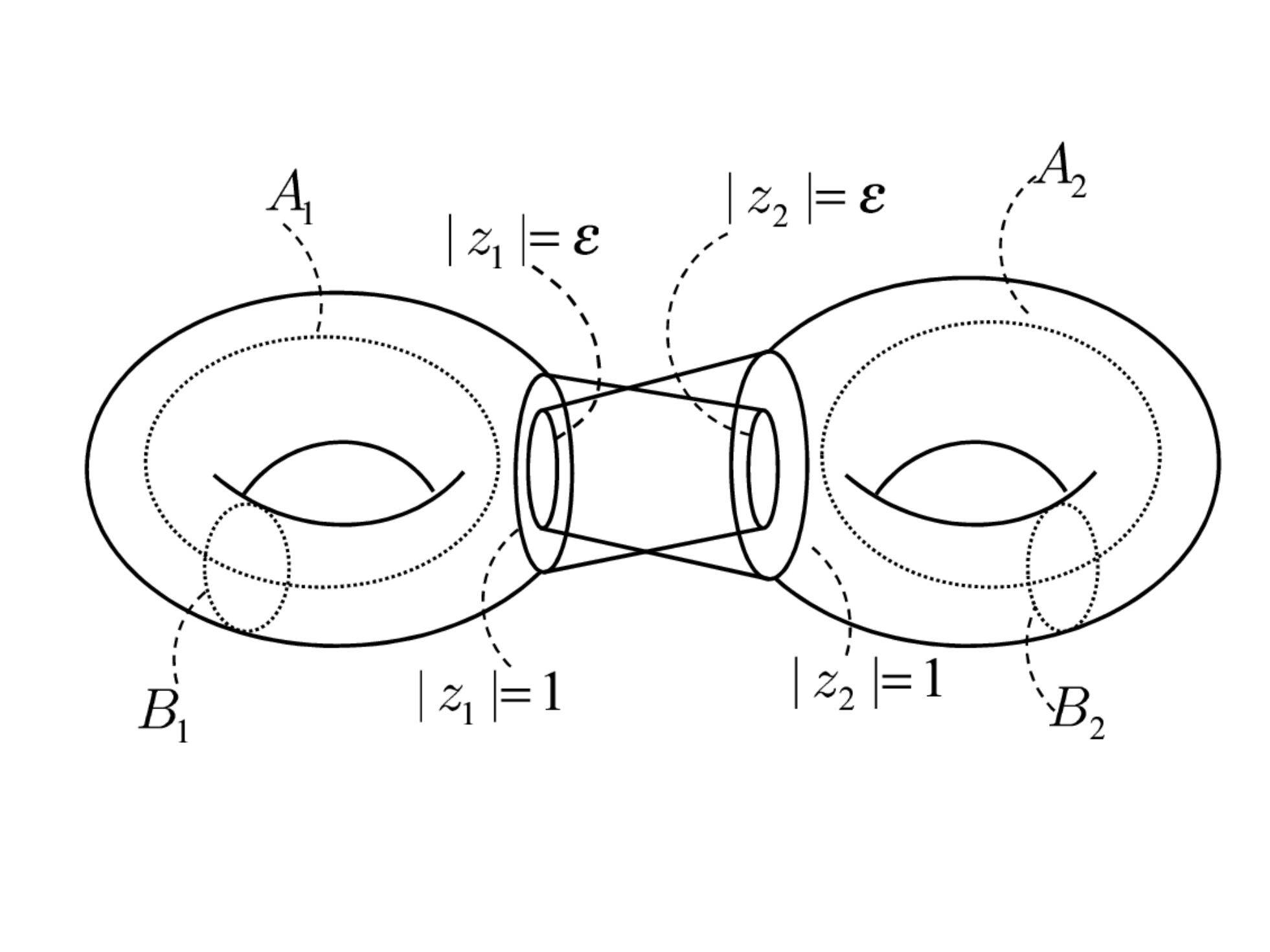}
\caption{Two tori are sewn together by identifying the annular regions 
$|\epsilon|\leq |z_a|\leq 1$ via the relation
$z_1 z_2 =\epsilon$. }
\label{fig:two-tori}
\end{figure}

In general, two compact Riemann surfaces $S_1$ and $S_2$ of genus $g_1$ and $g_2$ can be sewn together, giving a Riemann surface of genus $g_1+g_2$.
Choose complex local coordinates $z_a$ on $S_a$ ($a=1,2$), and excise the two disks $|z_a|<|\epsilon|$, where $\epsilon$ is a complex parameter satisfying $|\epsilon|<1$.
The centers of disks are taken at a point $z_a =0$.
The two surfaces are sewn together by identifying the annular regions 
$|\epsilon|\leq |z_a|\leq 1$ via the relation
\[z_1 z_2 =\epsilon.\]

By the way, for a pinching parameter $\epsilon$,
${\cal A}_a=\{\epsilon\leq |z_a|\leq 1\}\subset S_a$ is a cylinder to be identified.
Let ${\cal C}_a(z_a)\subset {\cal A}_a$ denote a simple closed, anti-clockwise oriented contour parametrized by $z_a$ surrounding the puncture at $z_a=0$, which is in relation of sewing ${\cal C}_1(z_1)\sim -{\cal C}_2(z_2)$ via $z_1z_2=\epsilon$. 

In the appendix, the derivations of $\omega(u,v), \nu_i, \varphi_{ij}$ are briefly shown.
The following is the HQD calculated in the context of pinching parameter \cite{YA}.
The HQD can be given by $\varphi_{ij}=\nu_i\nu_j$. $P_2(z)$ is a elliptic function related to the Weierstrass function and $E_n$ is Eisenstein series (their definition is given in the appendix),
\begin{align}
\varphi_{11}&=
\begin{cases}
\left(1+\epsilon^4 E_4(i)P_2''(i,z)\right)dz^2 +O(\epsilon^6) & z\in S_1 \\
 \epsilon^2 P_2(i,z)^2 dz^2 +6\epsilon^6 E_4(i)^2 P_2(i,z)^2 dz^2 +O(\epsilon^6)\in S_2 \\
\end{cases}\label{eqn:v11} \\
\varphi_{12}&=
\begin{cases}
-\epsilon P_2(i,z)dz^2+\epsilon^5\left(-3E_4(i)^2P_2(i,z)^2-\frac12E_4(i)P_2''(i,z)P_2(i,z) \right)dz^2+O(\epsilon^6)& z\in S_1 \\
-\epsilon P_2(i,z)dz^2 +\epsilon^5\left(-3E_4(i)^2P_2(i,z)^2-\frac12E_4(i)P_2''(i,z)P_2(i,z) \right)dz^2+O(\epsilon^6) & z\in S_2 \\
\end{cases} \label{eqn:v12}
\end{align}
and their permutation$(1\leftrightarrow 2)$.
Then we will illustrate their corresponding horizontal foliation in order to realize the deformation of the double torus.

\begin{figure}[htpb]
\centering
\includegraphics[width=15cm,clip]{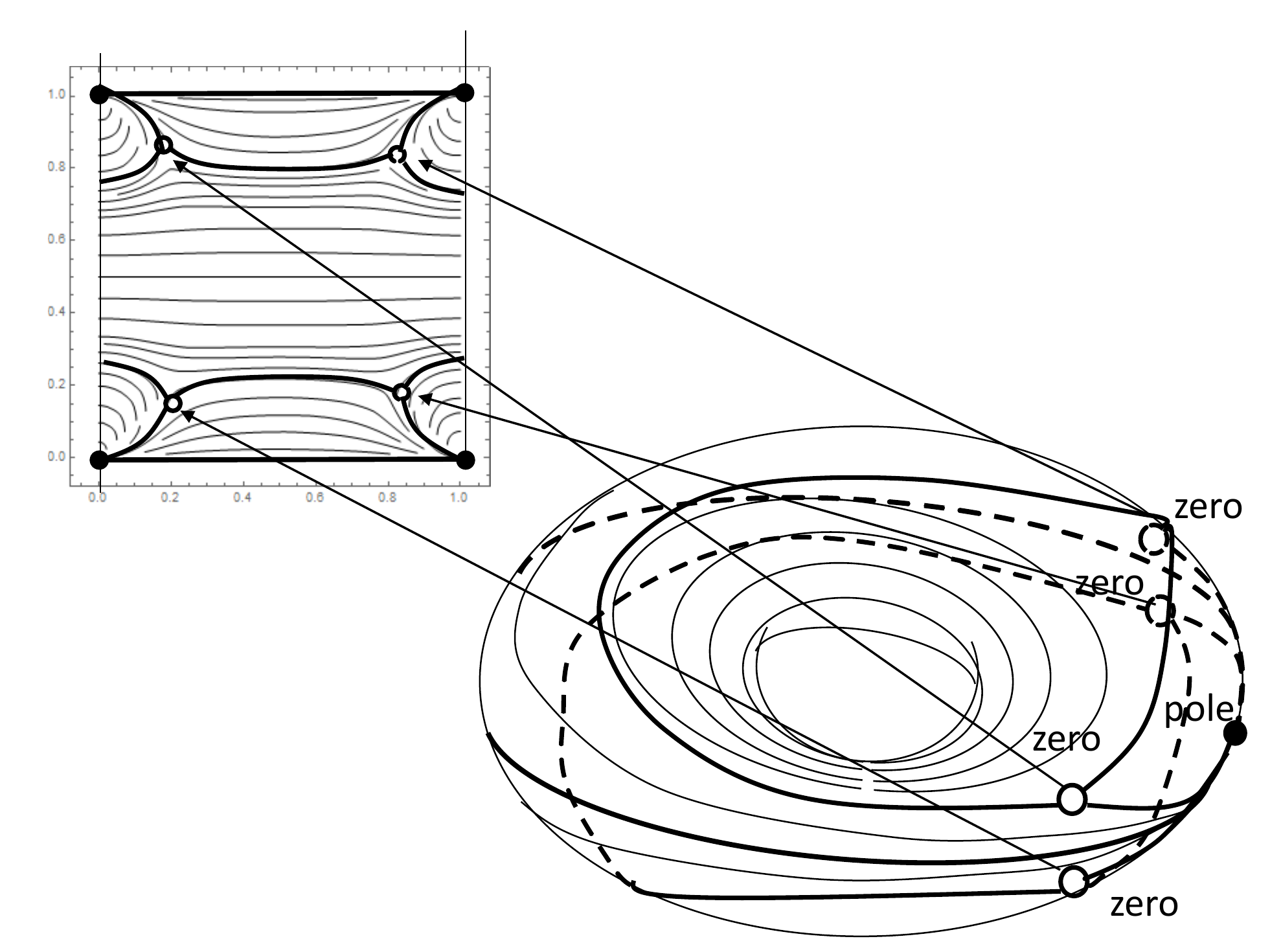}
\caption{The horizontal foliation of $\varphi_{11}(z\in S_1)$ is numerically calculated and shown on a fundamental square of the torus $S_1$, with Teichm\"{u}ller parameter $\tau_a=i$. There are one pole, and four trifurcations (of bold curves) for the zeros of order one.}
\label{fig:hf1}
\end{figure}

\begin{figure}[htpb]
\centering
\includegraphics[width=15cm,clip]{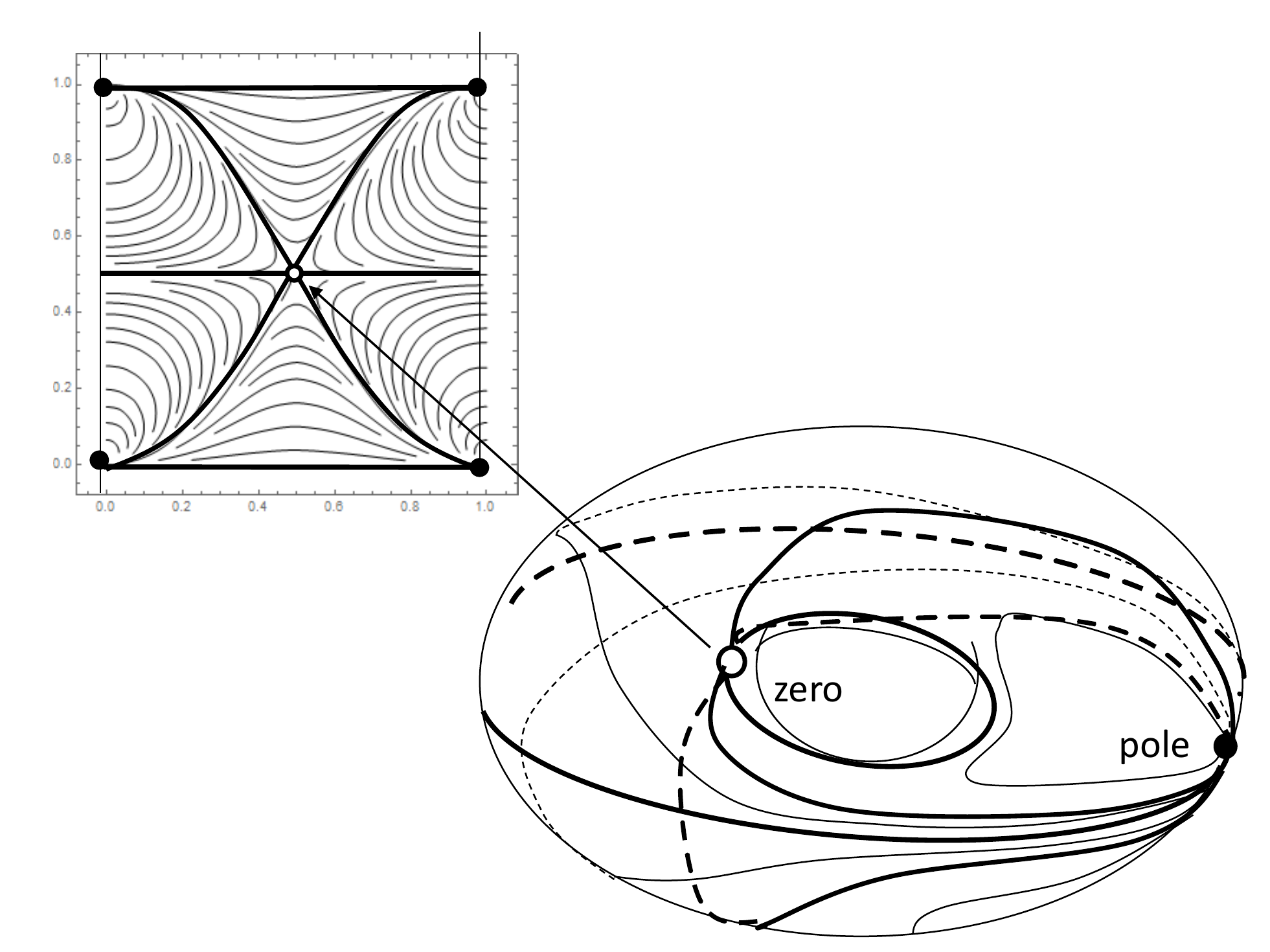}
\caption{The horizontal foliation of $\varphi_{11}(z\in S_2)$ is numerically calculated and shown on a fundamental square of the torus $S_2$, with Teichm\"{u}ller parameter $\tau_a=i$.  There are one pole, and one hexfurcation (of bold curves) for the zero of order four at the center. }
\label{fig:hf2}
\end{figure}

Firstly, we show the horizontal foliations for $\varphi_{11}, \varphi_{12}$. In Fig.\ref{fig:hf1} and Fig.\ref{fig:hf2}, the horizontal foliation of $\varphi_{11}(z\in S_1)$, $\varphi_{11}(z\in S_2)$ are numerically calculated and shown on a fundamental square of the tori $S_1$ and $S_2$, respectively, with Teichm\"{u}ller parameter $\tau_a=i$ as the base of background torus.
Moreover they are also illustrated on the tori $S_1$ and $S_2$.
The parameter $\epsilon$ is simply chosen to be $(0.001/E_4(i))^{1/4}$ so that $\varphi_{11}(z\in S_1)$ becomes $(1+0.001 P_2''(z))dz^2$.
On the other hand,  $\varphi_{11}(z\in S_2)$ is proportional to $P_2(z)^2 dz^2$.
In the figures, the pole (black point) and the zeros (white points) of $\varphi_{11}$ are shown.
All zeros in $S_1$ are of order one and the horizontal foliation has a trifurcation and its position is agree with the position of zeros which are numerically calculated as $\pm 0.196\pm 0.196 i+m+n i$, while in $S_2$ the zero at the center is of order four, so that it has a hexfurcation. 

\begin{figure}[htpb]
\centering
\includegraphics[width=15cm,clip]{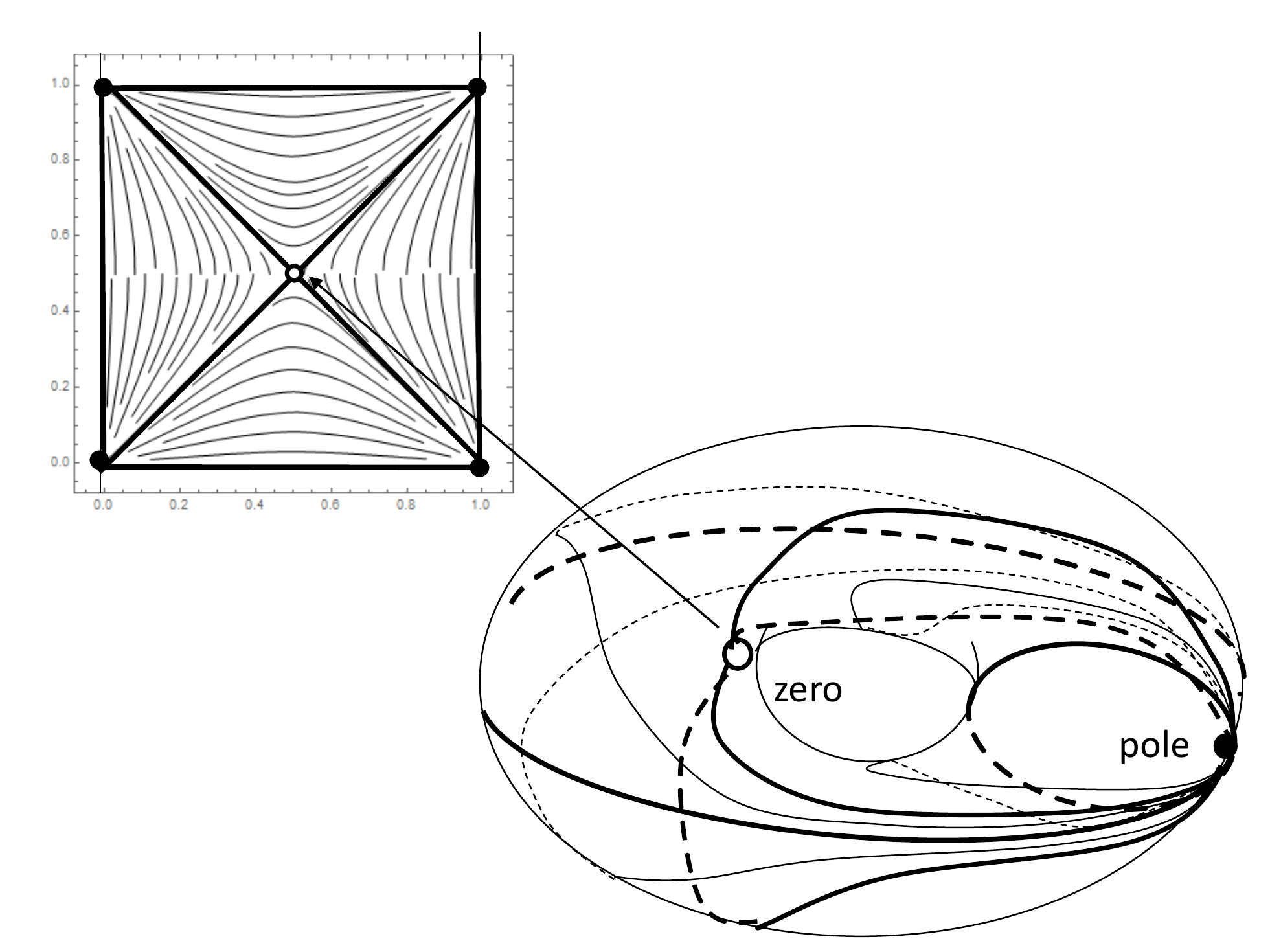}
\caption{The horizontal foliation of $\varphi_{12}(z\in S_2)$ is numerically calculated and shown on a fundamental square of the torus $S_1$, with Teichm\"{u}ller parameter $\tau_a=i$.  There are one pole, and one quadfurcation (of bold curves) for the zero of order two at the center.}
\label{fig:hf3}
\end{figure}

The horizontal foliation of $\varphi_{12}(z\in S_1)$ is shown in Fig.\ref{fig:hf3}.
$\epsilon$ is set to be $0.01$ and then 
$\varphi_{12}\propto P_2(z)+(0.01)^4(E_4(i)^2P_2(z)^2+0.5E_4(i) P_2(z)''P_2(z))$.
The zero of $\varphi_{12}$ is on the center of the square since $P_2(\tau=i,z)$ has a zero of order two at $z=0.5+0.5 i$. Then the horizontal foliation has a quadfurcation there.

Now it should be significantly noted that the expansion in $\epsilon$ becomes false near the zeros.
We have selected a not sufficiently small value $\epsilon$ for $\varphi_{11}$ in order to see typical configuration of the horizontal foliation.
Indeed the number of zeros of $\varphi_{11}$ with order is seemed to be incorrect, while HQD should have four zeros on a double torus.
Near the zeros the contribution of higher order becomes significant, so that some of zeros would approach to the pole and be canceled with the other structure (in $\epsilon=0$, $\varphi_{11}$ becomes constant and without zero as well as in a torus).
On the other hand, two degenerated zeros for $\varphi_{12}$ in a double torus might simply divide into four zeros of order one by the higher order contribution near zeros.

Except for such fine features of $\varphi_{11}$ and $\varphi_{12}$, their horizontal foliation will be representing the geometrical characteristics of `real' HQD.

\begin{figure}[htpb]
\centering
\includegraphics[width=15cm,clip]{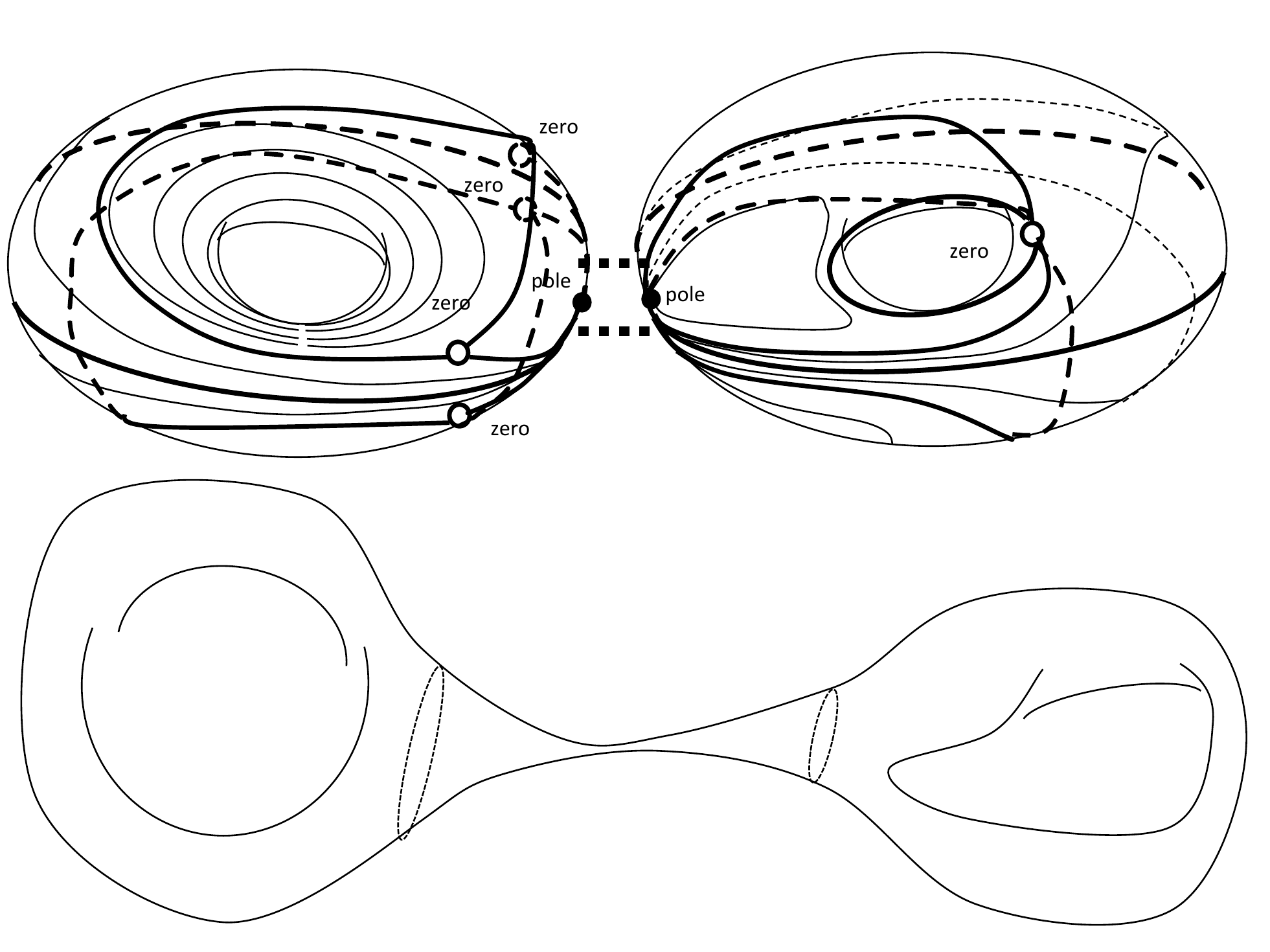}
\caption{The Teichm\"{u}ller deformation by the affine stretching of a double torus by $\varphi_{11}$ is illustrated. Since the HQD is approximately constant around the center, the torus $S_1$ is deformed like a Teichm\"{u}ller deformation of a single torus, there. }
\label{fig:hfd1}
\end{figure}

\begin{figure}[htpb]
\centering
\includegraphics[width=15cm,clip]{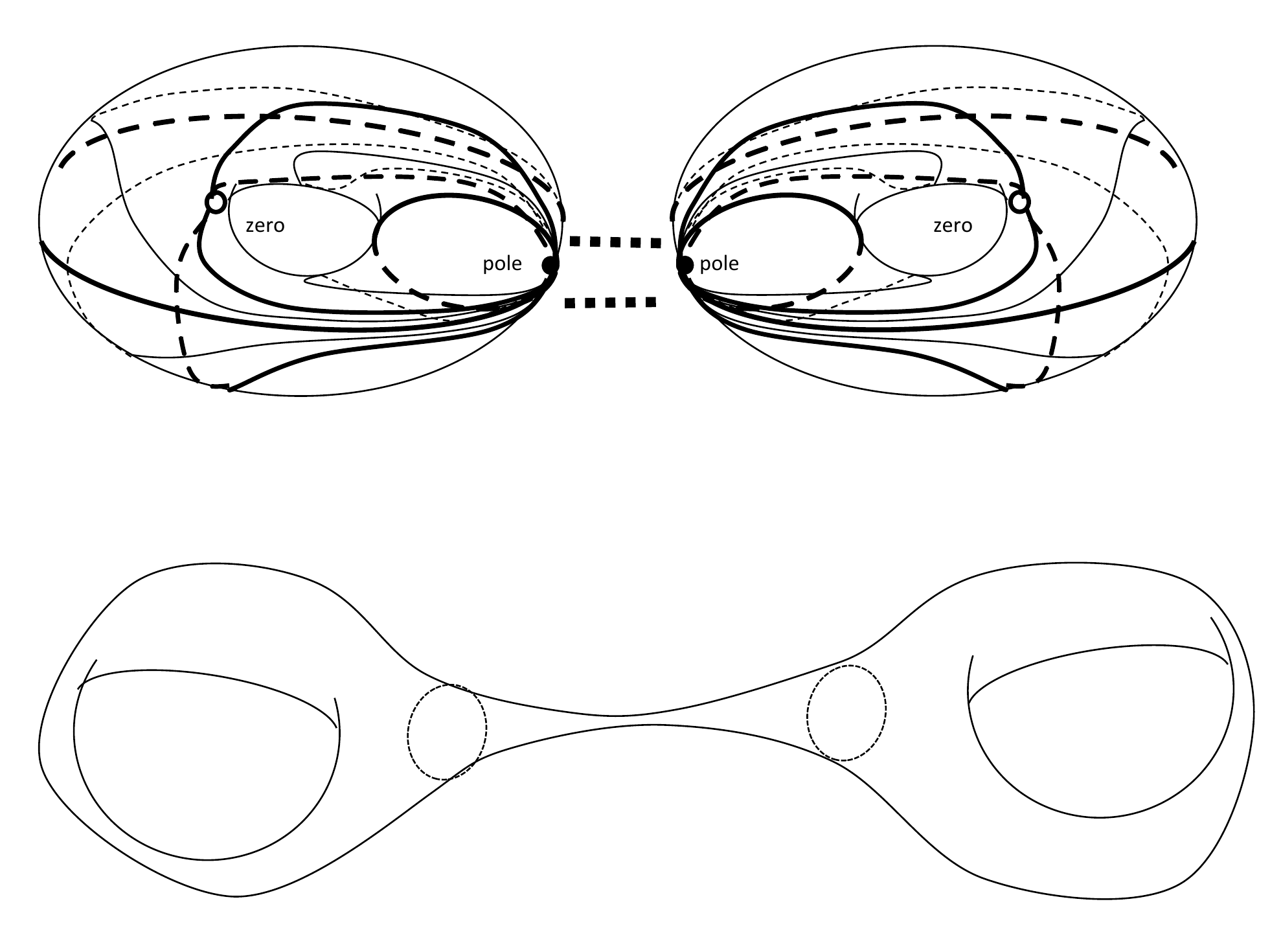}
\caption{The Teichm\"{u}ller deformation by the affine stretching of a double torus by $\varphi_{12}$ is illustrated. A torus $S_1$ and a torus $S_2$ are symmetric in this deformation. }
\label{fig:hfd2}
\end{figure}

In figures \ref{fig:hfd1} and \ref{fig:hfd2}, the Teichm\"{u}ller deformation by the affine stretching of a double torus by $\varphi_{11}$ and $\varphi_{12}$ are illustrated. By the affine stretching the Riemann surfaces are stretched in horizontal direction, and contracted in vertical direction. 
For $\varphi_{11}$, there seems to be the background contribution of Teichm\"{u}ller deformation of a torus around the center of the square.
For $\varphi_{12}$, we will see the symmetry about the changing two tori each other.

Since the Teichm\"{u}ller space is complex vector space $\langle \varphi_{ij}\rangle_{\mathbb C}$, there is another component $e^{i\theta}\varphi_{ij}$ of the deformation corresponding to rotation of the real axis. For example in the case of torus, this is depicted in Figure \ref{fig:hortor}. HQD for torus is constant $\varphi=\nu_r+i \nu_i \in {\mathbb C}$ and the rotation of the real axis provide a non-trivial changes of the horizontal foliation.
\footnote{These parameters are related to the conventional Teichm\"{u}ller parameter $\tau=\xi+i\eta$ as
\[
\xi=\frac{-2k\nu_R\nu_I}{(1-k)\nu_R^2+(1+k)\nu_I^2},\ \  \eta=\frac{\sqrt{1-k^2}(\nu_R^2+\nu_I^2)}{(1-k)\nu_R^2+(1+k)\nu_I^2}.
\]}

 Especially, rotation in right angle changes horizontal foliation to the vertical foliation.
In the affine stretching along the vertical foliation corresponds to a modular transformed $\tau\mapsto -\frac1{\tau}$ affine stretching.
In other sense, the vertical foliation gives the image of reverse deformation ($(1-k)$ contracted in real axis and $(1+k)$ stretched in imaginary axis).
We cannot deny other possibility of pattern not equivalent to these figures for large $\epsilon$.
To clarify that, we will need more systematic and exhaustive search.

\section{dynamics of Teichm\"{u}ller deformation of double torus}
Here we investigate the (2+1)-dimensional gravitational dynamics of the double torus universe. When once we admit the ADM-variables, the Einstein-Hilbert action gives definite evolution of the system.
\subsection{ADM-equations}

In the ADM-formalism\cite{JY}, the metric function is decomposed into
\begin{align}
ds^2=-N^2dt^2+h_{ab}(dx^a+N^adt)(dx^b+N^bdt),
\end{align}
where $N$ is a lapse function and $N_a$ is a shift vector.

Then the Einstein Hilbert action is given by
\begin{align}
S=\frac1{16\pi G}\int d^3x N\sqrt{h}\left[K^{ab}K_{ab}-trK^2+{}^{(2)}R\right]+S_m,
\label{eq:ac}
\end{align}
where $K_{ab}$ is extrinsic curvature, that is momentum variable of spatial geometry, 
${}^{(2)}R$ is two dimensional spatial scalar curvature, and $S_m$ is the contribution of the matter field.

As $N$ and $N_a$ are not dynamical variables but are gauge variables to be fixed, their consistency induces the constraint equations. Omitting the matter field for simplicity, the whole of   Einstein equations of ADM-formalism for vacuum spacetime with $\Lambda$ are
\begin{align}
&D_a ({K}^{ab}-h^{ab}trK)=0,\\
&K_{ab} K^{ab}-trK^2-{}^{(2)}R+2\Lambda=0,\\
&\frac{\partial}{\partial t}{K^{ab}}=\frac12Nh_{ab}(K^{ab}K_{ab}-trK^2-2\Lambda)+N(2K^{ac}K_c^b+K^{ab}trK)-D_aD_bN-{\cal L}_{N_a}K^{ab},
\end{align}
where $D_a$ is the covariant derivative with respect to the spacial metric $h_{ab}$ and ${\cal L}_{N_a}$ is the Lie derivative along the shift vector.
$K_{ab}$ is decomposed into its trace $tr K$ and a traceless part $\widetilde{K}_{ab}=K_{ab}-\frac12 trK h_{ab}$.
In our definition of $h_{ab}=\phi \hat{h}_{ab}, \ \ \ (|\hat{h}|=const.)$, their conjugate momentums are given by
\begin{align}
P_\phi&=\frac{\delta S}{\delta \phi}=tr K,\\
P_{\hat{g}}&=\frac{\delta S}{\delta \hat{h_{ab}}}=\widetilde{K}_{ab}.
\end{align}
If we consider simply a hypersurface with constant mean curvature \cite{HNM,SF}, the momentum constraints reduces to $D_a\widetilde{K}^{ab}=0$.
In comformally flat complex coordinate $(z,\bar{z})$, spatial metric is $h=\Omega^2dz d\bar{z}$, and then we know $\widetilde{K}_{zz}$ satisfying the momentum constraints should be a holomorphic quadratic differential, since the momentum constraints are reduced to $\partial_{\bar{z}} \widetilde{K}_{zz}=0$.

%On the other hand, as the canonical variable one may choose $h_{ab}$ and $\pi^{ab}=\sqrt{h}(K^{ab}-tr K h^{ab})$ then the constraint equations are given in

As the spatial hypersurface is homogeneous in our case, we simply take a constant trace of extrinsic curvature in consistent to comoving synchronous gauge.
As long as we are focus on the homogeneous spatial geometry, each term of the action functional (\ref{eq:ac}) is spatially constant and the spatial integral is reduced to the volume of the spatial section, which is
\begin{align}
&V(t)=a(t)^2 v(t)
=a(t)^2\int_R \sqrt{h_H}dzd\bar{z}\\&=a(t)^2\int_{R_0}\sqrt{h_{af}}d\xi d\bar{\xi}=a(t)^2\int_{R_0}\sqrt{h_H}\left|\frac{dz}{d\xi}\right||\varphi(z)|(1-k^2)d\xi d\bar{\xi},
\end{align}
where $a(t)$ is scale factor of spatial section introduced later and $h_H$ is the metric of the two dimensional hyperbolic space.

In ADM-formalism, the kinematic term of transverse traceless mode is given by the square of $\widetilde{K}_{ab}$ (that is the square about super-metric $G_{ijkl}$ for the canonical variable   $\pi^{ab}=\sqrt{h}(K^{ab}-tr K h^{ab})$). In terms of complex structure of Riemann surface that is the square norm of HQD about 
Weil-Petersson metric\cite{}. To define Weil-Petersson metric, firstly we define a Hermit inner product on a complex vector space $A_2(H,\Gamma)$ of HQD. The hyperbolic geometry $H$ is modeled by the Poincar\'{e} disk or the upper half plane.
For two HQDs $\phi=\phi(z)dz^2$ and $\psi=\psi(z)dz^2$,

\begin{align}
\langle \phi,\psi\rangle(z)=|h_H|^{-1}\phi(z)\overline{\psi(z)},\ \ \ (z\in H)
\end{align}
$\langle\phi,\psi\rangle$ is $\Gamma$ invariant and a function on a Riemann surface $R=H/\Gamma$, ($\Gamma$ is a discrete subgroup of $SO(2,1)\sim SL(2,R)$), where $ds_H^2=\sqrt{h_H}(z)|dz|^2$ is a metric of Poincar\'{e} disk $\sqrt{h_H}=\frac4{(1-|z|^2)^2}$ or Poincar\'{e} metric on the upper half space $\sqrt{h_H}=|\frac1{\Im z^2}|$.
Then we define a Hermit inner product of $\phi$ and $\psi$ as
\begin{align}
\langle\phi,\psi\rangle_R=&\iint_R\langle\phi,\psi\rangle d\sigma\\
=&\iint_R\sqrt{h_H}^{-1}\phi(z)\overline{\psi(z)}dxdy
\label{eq:pi}
\end{align}
This is the Petersson inner product of $A_2(H,\Gamma)$ by the Weil-Petersson metric and makes it a Hilbert space.
In $(z,\bar{z})$ coordinate, considering $\phi(z)=\widetilde{K}_{zz},\ \ \ \psi(z)=\widetilde{K'}_{zz}$,
\begin{align}
=&\iint_R\sqrt{h_H}^{-1}\widetilde{K}_{zz}\widetilde{K'}_{\bar{z}\bar{z}}dxdy \\
=&\iint_R\sqrt{h_H}\widetilde{K}_{zz}\widetilde{K'}_{\bar{z}\bar{z}}h^{z\bar{z}}h^{z\bar{z}}dxdy.
\end{align}
Then one will notice that this inner product gives the kinematic term of transverse traceless mode of
$\widetilde{K}_{ab}$ in Einstein-Hilbert action.
Nevertheless as the HQD and metric function of the Poincar\'{e} disk (or upper half plane) is not constant, the region of integration in (\ref{eq:pi}) which is a Riemann surface realized on the hyperbolic space $H$, also contains the information about Teichm\"{u}ller parameter.

To avoid this difficulty, rather we use a standard homogeneous metric constructed in \cite{MS1} by affine stretching.
Since the standard metric represent a Riemann surface realized on Poincar\'{e} disk metric by a standard fundamental region with a metric deformed by the pseudo-conformal mapping related to the affine stretching. As a base of the standard metric we always consider the torus with Teichm\"{u}ller parameter $\tau=i$.
Thus it is more simple to give kinematic term by calculating the Einstein-Hilbert action of the standard metric directly on the standard region which is independent of the parameter of Riemann surface $R$ than by analyzing from the Weil-Petersson inner product on the Riemann surface $R$. 

%$\phi$ determines harmonic Beltrami differential 

%\begin{align}
%\mu[\phi](z)=\lambda_H(z)^{-2}\overline{\phi(z)},\ \ \ (z\in H)\\
%H[\mu](z)=
%\end{align}

%Considering a holomorphic tangent space $T_0(T(\Gamma))$ of the Teichm\''{u}ller space $T(\Gamma)$, ( $T_0(T(\Gamma))\simeq B(H,\Gamma)/N(\Gamma)\simeq HB(H,\Gamma)$
%For $\mu_1,\mu_2\in B(H,\Gamma)$,

%\begin{align}
%h(\mu_1,\mu_2)=\iint_F\lambda_H(z)^2\mu_1(z)\overline{\mu_2(z)}dxdy
%\end{align}
%Then that is related to the Petersson inner product

%\begin{align}
%h(H[\mu_1],H[\mu_2])=\langle\phi[\mu_2]\phi[\mu_1]\rangle_R,\ \ \ (\mu\in B(H,\Gamma))
%\end{align}

%We identify $T_p(T(\Gamma))$ and $HB(H,\Gamma^\nu)$

About the Weil-Petersson metric, many important knowledge are explored, e.g., it is incomplete, K\"{a}hler, and with negative sectional curvature bounded above by a certain constant.
Nevertheless, in our purpose direct calculation seemed simplest way to have kinetic term of Einstein-Hilbert action.

\subsection{gravitational action for double torus}

%Firstly, we calculate the Petersson inner product of the HQD given in our previous work\cite{MS}.
%\begin{align}
%\varphi_{11}=
%\end{align}

%Since $\langle\cdot,\cdot\rangle_{\Delta}=\langle\cdot,\cdot\rangle_{H}$

%\begin{align}
%\langle\varphi_{11},\varphi_{11}\rangle=\\
%\langle\varphi_{11},\varphi_{12}\rangle=\\
%\langle\varphi_{12},\varphi_{12}\rangle=
%\end{align}

%From these coeficients, we have a new basis of $A_R$  which is orthonormal.

%\[\Phi_{11}, \Phi_{12},\Phi_{22}\]

%On the other hand, it was known that the WP-metric admit complex structure furthermore
%the HQDs are orthogonal.

%Then if asign $P_{ij}=\frac{}{}$, 
%The WP metric is 
%written as 

%\begin{align}
%%\end{align}

%The evolution is resolved by knowing the geodesics in the Poinc\'{a}re metric which
%is the half circle whose center is on the $\xi_I$ axis and the vertical line.

%Moreover if there is the contribution from the conforaml part.
%It changes the relation between the affine length of the geodesics and the coordinate time.
%That is easily understood by solving the followin equation.

%Then the dynamics can terminate in that upper half space.

Though the Weil-Petersson metric is not concretely known in an appropriate coordinate of Teichm\"{u}ller space, we will calculate all from the homogeneous standard metric of the Riemann surface.

As that allows us to determine the kinematic term of ADM-variables, the result corresponds to the concrete evaluation of the Weil-Petersson metric.
Moreover, since the spatial section is homogeneous, we only need to evaluate the action at a point in the homogeneous spatial section. Then spatial integration reduces to the volume. 
(If considering $\varphi_{ij}$ is rigorous holomorphic quadratic differential, the homogeneity is absolutely assured. It should be noted that actually, $\varphi_{ij}$ we handle here, however, is approximated in $\epsilon$-expansion and its homogeneity is assured only in certain order of $O(\epsilon^n)$. )

In terms of the HQD and the affine stretching, the standard metric is given by\cite{MS1},
\begin{align}
dl^2=h^S_{ab}dx^a dx^b\frac{4 a(t)^2}{(1-|z|^2)^2}\left|\frac{d z}{d\xi}\right||\varphi||d\xi+k\frac{\overline{\varphi}}{|\varphi|}d\bar{\xi}|^2,
\label{eq:hab}
\end{align}
where HQD $\varphi$ is a linear combination of $\varphi_I$, $\varphi=Q^I(t)\varphi_I$.
$\left|\frac{d z}{d\xi}\right|$ is Jacobian for conformal embedding of a square into the Poincar\'{e} disk and carries no information of the Riemann surface.

%Nevertheless we had better take a Beltrami coefficients as a parameter of dynamical deformation,
% for as coordinates of Teichmuller space, 
% $B(H,\Gamma)/N(\Gamma)$ is isomorphic to the dual space of $A_2(H,\Gamma=R)$
% by linear functional $\Lambda_{\mu}$
% \begin{align}
% \Lambda_\mu(\varphi)=(\mu,\phi)_R=\int\int_F\mu(z)\varphi(z)dxdy
% \end{align}
% where F is the fundamental region of $\Gamma$
%It is natural that the Beltrami coefficents are coordinate of configurationa space
%if $\tilde K_ij$  is HQD.

%Since also  $B(H,\Gamma)$ is  linear space, 
%$\mu_f=k(t) $

Here the spatial metric is given by the pull back of a trivial metric by a coordinate transformation.
Thinking of time-dependence of the coordinate transformation, a gauge fixed spacetime metric is not a trivial spacetime metric, but a dynamical one, for example $ds^2=-N^2dt^2+h^S_{ab}(t)dx^a dx^b$.
Now we factorize the spatial metric $h^S(t)=\Omega^2\hat{h}^S(t)$, where $|\hat{h}^S|$ is constant.
$N=1$ is consistent to the homogeneous spatial hypersurface.

\begin{align}
&\hat{h}^S_{ab}(t)dx^a dx^b=|d\xi+kFd\bar{\xi}|^2,\ \ \ F=\frac{\overline{\varphi}(\xi)}{|\varphi(\xi)|} , \ \ \ |\hat{h}^S|=1-k^2 ,\\
& |\Omega|^2=\frac4{(1-|z|^2)}\left|\frac{dz}{d\xi}\right||\varphi(\xi)|,\\
&tr K=-2\frac{\dot{\Omega}}{\Omega} \\
&\widetilde{K}_{ab}dx^adx^b=-\frac12\Omega^2\dot{\hat{h}}^S_{ab}dx^adx^b=-\frac12\Omega^2
(k\dot{F}d\xi^2+k\dot{\overline{F}}d\bar{\xi}^2). \label{eq:ttK}
\end{align}

From $F \overline{F}=1$, we have
\begin{align}
&\widetilde{K}_{ab}\widetilde{K}^{ab}= \frac{2k^2\dot{F}\dot{\overline{F}}}{(1-k^2)^2}=-\frac{2k^2\dot{F}^2}{(1-k^2)^2F^2}=-\frac{2k^2(\frac{\partial}{\partial t} \log{F})^2}{(1-k^2)^2}.
\end{align}
Furthermore the kinematic term reduces to
\begin{align}
&\log F=\log\frac{\overline{\varphi}}{|\varphi|}=\log\frac{\overline{Q^I(t)\varphi_I(\xi)}}{\sqrt{Q^I(t)\varphi_I(\xi)\overline{Q^I(t)\varphi_I(\xi)}}}
=\frac12\log \overline{Q^I(t)\varphi_I(\xi)}-\frac12\log Q^I\varphi_I(\xi)\\
&\frac{\partial}{\partial t} \log{F}=\frac{\overline{\dot{Q^I}\varphi_I}}{2\overline{Q^I\varphi_I}}-\frac{ \dot{Q}^I\varphi_I}{2 Q^I\varphi_I}\\
&\widetilde{K}_{ab}\widetilde{K}^{ab}=\frac{2k^2}{(1-k^2)^2}\left[\Im\left(\frac{\dot{Q^I\varphi}_I}{Q^I\varphi_I}\right)\right]^2.
\end{align}

Since the trace less part of extrinsic curvature $\widetilde{K}_{ab}$ is HQD $\varphi$ and momentum constraint is satisfied.

Nevertheless in this parametrization, all the degrees of freedom are not incorporated since $\widetilde{K}_{ab}$ in (\ref{eq:ttK}) is same for parallel HQDs $\varphi'=c \varphi,\ \ \ c\in{\mathbb R}$.
This redundancy are compensated in the $tr K$ as,
\begin{align}
(tr K)^2&=\left(\frac{\partial}{\partial t}(\log \Omega)\right)^2
=\left(\frac{\dot{a}}{a}+\dot{\log|\varphi|}\right)^2
=\left(\frac{\dot{a}}{a}+\dot{\frac12\log \varphi\overline{\varphi}}\right)^2 \\
&=\left(\frac{\dot{a}}{a}+\frac12\frac{\dot{Q^I}\varphi_I}{Q^I\varphi_I}+\frac12\frac{\overline{\dot{Q^I}\varphi_I}}{\overline{Q^I\varphi_I}}\right)^2=\left(\frac{\dot{a}}{a}+\Re\left(\frac{\dot{Q}^I\varphi_I}{Q^I\varphi_I}\right) \right)^2\\
&S=\int dt a(t)^2 v(t) \left[\ \ \ \right]_{\xi=\xi_0},
\end{align}
where $\xi_0$ should not be the zero of the HQDs.
%This fact essentially gives curvature to the Weil-Petersson metric.
Though that gives the coupled kinematic term, of course, we can get a decoupled kinematic term by rearranging the scale factor $a(t)$.
%It only gives Weil-Petersson metric in a special coordinate and new variable.
Complicated structure of Weil-Petersson metric is embedded into the values of $\varphi(\xi_0)$ which can be unpacked by the numerical value of Weierstrass elliptic function.
Nevertheless, there will be not much profit even if we analyze it in detail, now. 
Rather we are concentrate on a limited initial condition.

In limiting initial degrees of freedom on the direction along one HQD, eternally the dynamics is limited there, since the system is isotropic for basis $\{\varphi_I\}$ of complex three dimensional vector space $A_2(R)$, while it is not so in the complex plane of $\varphi(\xi)\in \mathbb{C}$.
In the next subsection, we attempt to analyze the dynamics, in the limited degrees of freedom.

.

\subsection{dynamics for a single HQD}

Here we limit the degrees of freedom on the two dimensions (one dimension in complex variable) along one HQD $\varphi_L$.
From the Teichm\"{u}ller theorem \cite{A94}, for $\varphi\in A_2(R)_1$ we can regard a Beltrami coefficient\cite{HN2}\cite{IT} as a representation of Teichm\"{u}ller space by choosing
$k=||\varphi||_1$ in equation (\ref{eq:hab}) as
\begin{align}
\mu_{\phi}=||\phi||_1\frac{\overline{\varphi(\xi)}}{|\varphi(\xi)|}.
\end{align}

Now appointing $\varphi_L^1=\varphi_L/||\varphi_L||_1$ to a basis, we assign the HQD parallel to $\varphi_L$ as
\begin{align}
\varphi=(Q_r(t)+i Q_i(t))\varphi_L^1,\ \ \ \ (|Q=Q_r+i Q_i|<1, \ \ \ Q_r, Q_i\in \mathbb{R}).
\end{align}
then $||\varphi||_1=|Q| ||\varphi_L^1||=|Q|$.

Consequently the spatial metric becomes
\begin{align}
dl^2=h^S_abdx^adx^b=|\Omega|^2 \left|d\xi+k \frac{\overline{\varphi(\xi)}}{|\varphi(\xi)|}d\bar{\xi}\right|^2
=|\Omega|^2 \left|d\xi+(Q_r(t)+i Q_i(t)) \frac{\overline{\varphi_L^1(\xi)}}{|\varphi_L^1(\xi)|}d\bar{\xi}\right|^2,
\end{align}
where $|\Omega|^2=a(t)^2\frac{4|dz/d\xi|^2}{(1-|z|^2)^2}|\varphi|=a(t)^2\frac{4|dz/d\xi|^2}{(1-|z|^2)^2}|Q(t)||\phi_L^1(\xi)|$, for  $|\varphi|=|P||\phi_L^1|$. $v_0$ is constant.

For a convenience in the following, we redefine $A(t)^2=a(t)^2|Q(t)|(1-|Q(t)|^2)$, so that
$V(t)=a(t)^2 v(t)=a(t)^2\int_R \sqrt{h_H}=a(t)^2\int_{R_0}\sqrt{h_S}=a(t)^2\int_{R_0}\sqrt{h_H}|\varphi||1-|Q|^2|=A(t)^2\int_{R_0}\sqrt{h_H}|\varphi_L^1|=A(t)^2v_0$

\begin{align}
dl^2=\frac{4A(t)^2|\frac{dz}{d\xi}|^2}{(1-|z|^2)^2}\frac{|\varphi_L^1(\xi)|}{1-|Q(t)|^2}\left|d\xi+(Q_r(t)+i Q_i(t)) \frac{\overline{\varphi_L^1(\xi)}}{|\varphi_L^1(\xi)|}d\bar{\xi}\right|^2
\end{align}

Since this is homogeneous spatial metric, we can choose $N=A(t)^4$ and $N^a=0$;
\begin{align}
ds^2=-A(t)^4dt^2+\frac{4A(t)^2|\frac{dz}{d\xi}|^2}{(1-|z|^2)^2}\frac{|\varphi_L^1(\xi)|}{1-|Q(t)|^2}\left|d\xi+(Q_r(t)+i Q_i(t)) \frac{\overline{\varphi_L^1(\xi)}}{|\varphi_L^1(\xi)|}d\bar{\xi}\right|^2.
\end{align}
Then calculating scalar curvature directly, the Einstein Hilbert action becomes

\begin{align}
S=\int_{R_0}(R-2\Lambda) \sqrt{g}dx^3=v_0\int dt\left[ -6\left(\frac{\dot{A}}{A}\right)^2+4\frac{\ddot{A}}{A}+\frac{2(\dot{Q_r}^2+\dot{Q_i}^2)}{(1-|Q|^2)^2}-2A(t)^2|Q|(1-|Q|^2)-2\Lambda A(t)^4\right] \ .
\end{align}

The fourth term is coming from spatial Ricci scalar curvature.
One may see that this system is similar to a particle moving in the two dimensional hyperbolic space.
If there is no coupling with scale factor and the Teichm\"{u}ller parameter, the solution is the geodesic of Poincar\'{e} disk which is circles perpendicular to the circumference at infinity $|Q|=1$.

As we have a coupling through spatial curvature 
\begin{align}
{}^{(2)}R=2 \frac{|Q|(1-|Q|^2)}{A(t)^2}
\end{align}

This coupling gives a force along the radial direction for $\vec{Q}=(Q_r, Q_i)$.
Similarly, $A(t)$ is affected by a coupling force also.
Their equations of motion are given by
\begin{align}
& (1-|Q|^2)\ddot{Q}_r+2\dot{Q}_r^2Q_r+4\dot{Q}_r\dot{Q}_iQ_i-2\dot{Q}_i^2Q_r=- \frac{A^2Q_r(1-|Q|^2)^3(1-3Q_r^2-3Q_i^2)}{|Q|}  \label{eq:eom1}\\
& (1-|Q|^2)\ddot{Q}_i+2\dot{Q}_i^2Q_i+4\dot{Q}_i\dot{Q}_rQ_r-2\dot{Q}_r^2Q_i= - \frac{A^2Q_i(1-|Q|^2)^3(1-3Q_r^2-3Q_i^2)}{|Q|}   \label{eq:eom2}\\
& \ddot{A}-\dot{A}^2/A-2\Lambda A^5= A(1-|Q|^2)|Q|.\label{eq:eom3}
\end{align}

\begin{figure}[htpb]
\centering
\includegraphics[width=15cm,clip]{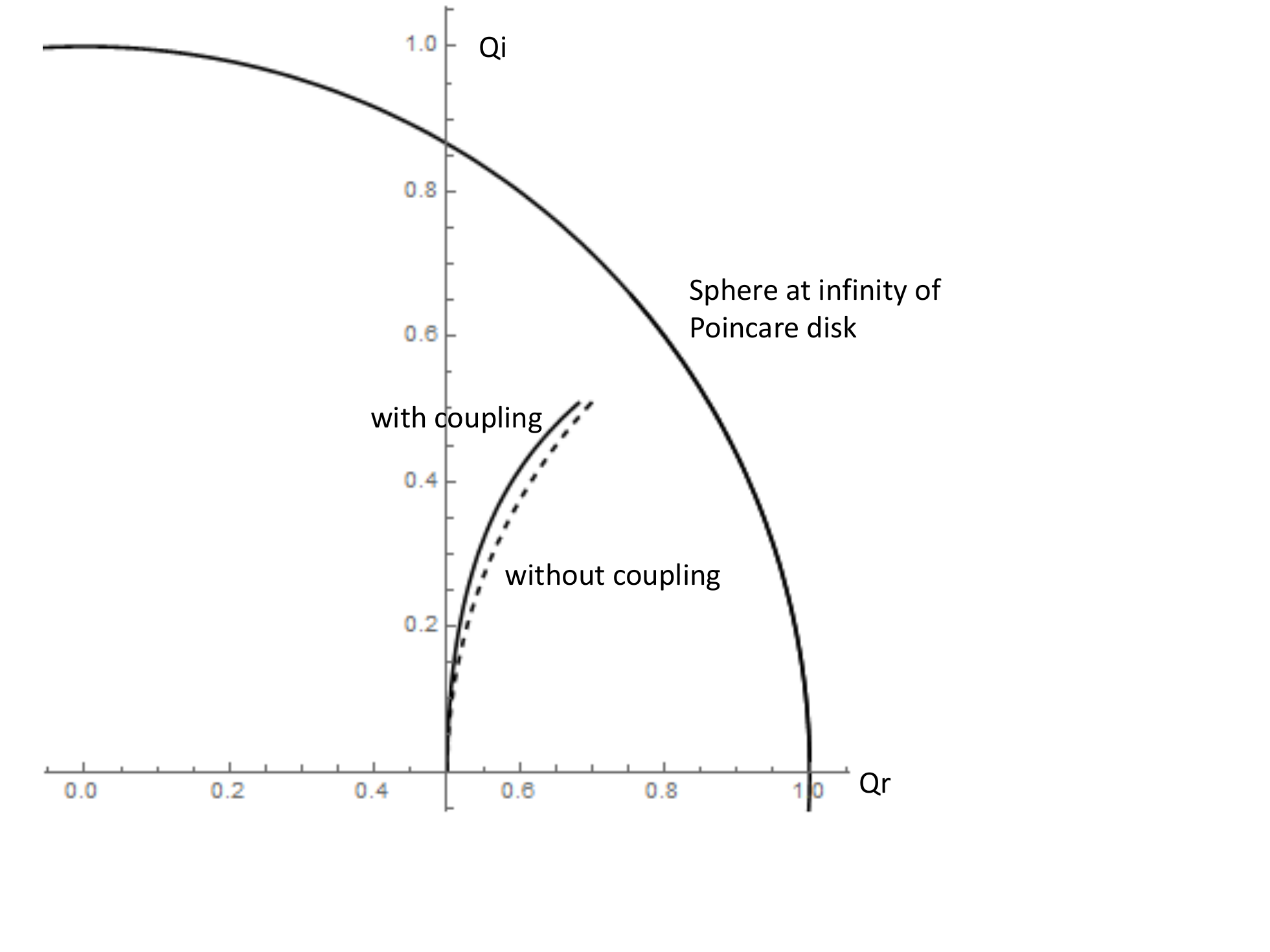}
\caption{The large circle is a circumference at infinity of the Poincar\'{e} disk. The orbit of $\vec{Q}=(Q_r,Q_i)$ without coupling are drawn in a broken curve and is on the geodesic of the Poincar\'{e} disk, which is a circle perpendicular to the circumference at infinity. 
The orbit of real solution with coupling is drawn in a solid curve, and deformed from the geodesic by   central force for the coupling. These curves are incomplete because of the divergence of $A(t)$.}
\label{fig:gr-1}
\end{figure}

As an example, we numerically solve the equation (\ref{eq:eom1}), (\ref{eq:eom2}), (\ref{eq:eom3}),
with initial conditions $Q_r(0)=0.5, \ \dot{Q}_r(0)=0,\ Q_i(0)=0,\ \dot{Q}_i(0)=0.5,\ A(0)=1,\ \dot{A}(0)=0$ and $\Lambda=0.25$.
Fig.\ref{fig:gr-1} shows the orbit of $\vec{Q}=(Q_r,Q_i)$. A broken curve is the solution omitting the interaction of RHS in (\ref{eq:eom1}), (\ref{eq:eom2}). As expected, it is on the geodesic of Poincar\'{e} disk which is a circle crossing the circumference at infinity $|Q|=1$ in right angle.
The solid curve is the real solution with interaction. We see that the coupling force deform the trajectory from the geodesic of Poincar\'{e} disk metric.
The interaction term only depend $|Q|$ and $A$, $\vec{Q}=(Q_r,Q_i)$ seems to be affected by an inward central force.
Both curves are not complete since they do not reach to the circumference at infinity.
The endpoints are caused by the divergence of $A(t)$, which is similarly observed in toroidal case \cite{SF}.

Understanding this dynamics from the viewpoint of affine stretching, at $t=0$ the Riemann surface is affine stretched by a real coefficients $Q_0=0.5+ 0.0 i$ of $Q_0\varphi_L \in A_2(R)$, as in figures\ref{fig:hfd1}, \ref{fig:hfd2} along each horizontal foliation of $\varphi_L$.
On $t>0$ imaginary part of coefficients increase and $|Q|$ also increase. Then the Riemann surface is further stretched, but its direction of the affine stretching is rotated in $\arg(Q_r+iQ_i)$. 
The corresponding horizontal foliation is rotated one similar to Fig.\ref{fig:hortor} of toroidal case. For example 
in choosing $\varphi_{11}$, the torus of $S_1$ becomes slim and increasing its Dehn twist\cite{}.
The opposite torus of $S_2$ will be squeezed so complicatedly.

\begin{figure}[htpb]
\centering
\includegraphics[width=15cm,clip]{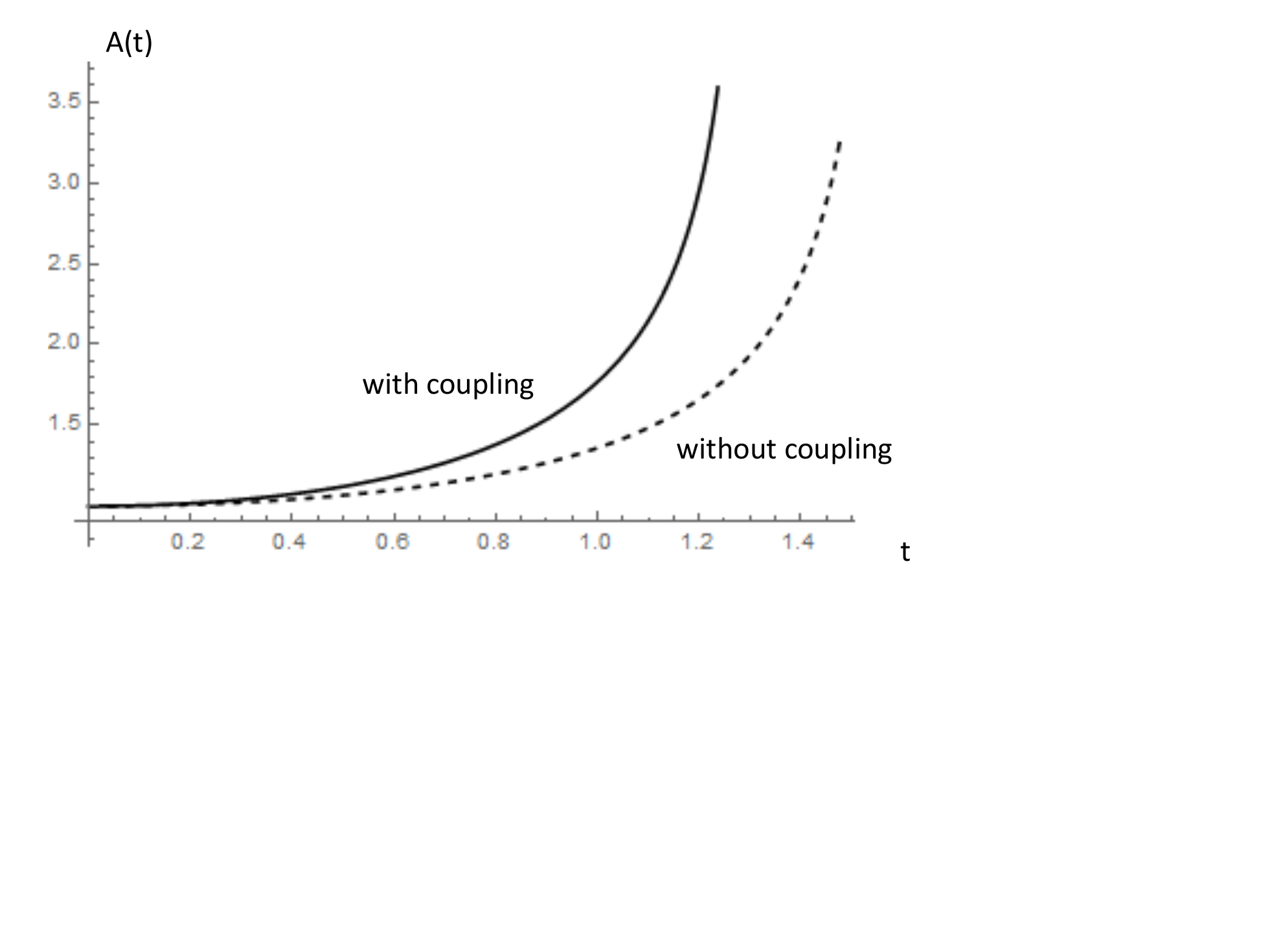}
\caption{The evolution of $A(t)$ are shown both in the case of with coupling (solid curve) and without coupling (broken curve). They are divergent in finite time, since the time coordinate is rescaled by $N=A(t)^4$.}
\label{fig:gr-2}
\end{figure}

Fig. \ref{fig:gr-2} shows the evolution of $A(t)$ which is related to the volume of the double torus. 
With positive $\Lambda=0.25$, both $A(t)$ without coupling and with coupling, inflationary grow and they are divergent in finite time, since the time coordinate is rescaled by $N=A(t)^4$. The expansion of the universe is enhanced by the interaction term, as well as by simple negative curvature term.

\subsection{product-spacetime}
Thurstone's geometrization conjecture has been confirmed by Perelman\cite{P}.
In the investigation of gravitational dynamics of the universe, these geometrical models will become next candidates of the globally anisotropic universe to homogeneous isotropic FRW-model.

The eight geometrical models are listed in the table.
As they are related to the Bianchi type homogeneous space ($S^2\times {\bf R}$ will be related to Kantowski-Sachs type), the local dynamics has been already formulated, and in some cases, concretely solved. 
Nevertheless, since we consider the compact manifold as a cosmological model, also global degrees of freedom should be handled.
Among the eight geometrical models, they were well studied except for the $H^2$-fibration space. Now on, the present standard metric formulation also will be admitted in a direct product of $H^2\times X$ as we have succeeded it on $H^2$, which was the last  peace of such anisotropic cosmological models.

\begin{table}[htb]
  \begin{tabular}{|c|c|c|} \hline
    geometrical model & structure  &   Bianchi type \\ \hline \hline
   ${ \bf R^3}$ & isotropic & I \\
  $ {\bf  S^3}$ & isotropic & IX \\
 $  {\mathbb H}^3$ & isotropic & V \\ 
   $ S^2\times R$ & fibre on $S^2$ & no \\
   $ H^2\times R$ & fibre on $H^2$ & III \\
    Nil & fibre on $R^2$ &  II \\
    $ \tilde{SL(2,R)}$ & fibre on $H^2$ & VIII \\
    Sol & fibre on $R$ & VI${}_0$\\ \hline
  \end{tabular}
  \label{tab:1}
\end{table}

\section{summary and discussions}
We have investigated the dynamics of double torus universe with locally hyperbolic geometry for its degrees of freedom for Teichm\''{u}ller deformation.
We handled the finite degrees of freedom incorporating into the local homogeneous standard metric so that automatically Weil-Petersson geometry is considered.
The deformation modes are illustrated in a context of horizontal foliation.
Especially, in the case of the dynamics is limited along one holomorphic quadratic differential, evolution of the Teichm\"{u}ller deformation is solved.
As suggested in \cite{HN2}, in order to treat Hamiltonian formalism for canonical quantization, the Beltrami coefficient may be treated as a coordinate of configuration space as in the case of a single HQD.

Moreover, especially limiting the motion on a two dimensional section of the Teichm\"{u}ller space along a holomorphic quadratic differential, we have solved the dynamical motion of the degrees of freedom associated with the Teichm\"{u}ller deformation.
It was shown that the trajectory in configuration space is changed from the geodesic of 
Poincar\'{e} disk metric by the coupling force.

Though our analysis is only in leading order of the pinching parameter, in principle all regions of the configuration space can be covered by the full order calculation of the pinching parameter.
Moreover while our concrete solution have been limited in a section of phase space along one HQD, lengthy calculation for the Weil-Petersson metric and numerical integration of the equations of motion would give general solutions.
It is known that the Weil-Petersson metric is incomplete, K\"{a}hler, and with negative sectional curvature bounded above by a certain constant.
With the calculation along the present investigation, one might analyze asymptotic structure around such incomplete points, since the pinching parameter is introduced to approximate the Teichm\"{u}ller space around there.
Especially, for certain values of Teichm\"{u}ller parameters we expect there are any inherent additional structure.
Indeed, in our analysis some Eisenstein serieses inevitably vanish, as we have chosen special Teichm\"{u}ller parameters as a base point of the Teichm\"{u}ller deformation by the affine stretching, which would be the result of the existence of complex multiplication\cite{CM}. 
Related to the theory of the complex multiplication of the elliptic function, it might be revealed that any additional asymptotic symmetry of the Weil-Petersson metric appears around such a special incomplete point.

\section*{Acknowledgements}
We would like to thank to Prof. H. Miyaji for his helpful advices about mathematics of Riemann surface.

\section*{Appendix. holomorphic quadratic differential}

The following is formulae to determine differential forms $\omega(u,v), \nu_i, \varphi_{ij}$ in pinching double torus, that is sewn two tori $S_a \ \ (a=1,2)$\cite{YA}.
The unique bilinear two form $\omega(u,v)$ is given by
\begin{align}
\omega(u,v)=
\begin{cases}
\omega^{(a)}(u,v)+a_a(u)X_{\bar{a}\bar{a}}a_a^T(v) & u,v\in S_a \\
a_a(u)(-I+X_{\bar{a}a})a_{\bar{a}}^T(v) & u\in S_a, v\in S_{\bar{a}} \\
\end{cases}
\end{align}
where $\omega^{(a)}(u,v)$ is a unique bilinear two-form of $S_a$, $a_a(u)[k]$ is row vector
and $X_{ab}[k,l]$ is matrix indexed by $k,l=1,2,...$, defined by
\begin{align}
X_{ab}[k,l]&=\frac{\epsilon^{(k+l)/2}}{\sqrt{kl}}\frac1{(2\pi i)^2}\oint_{{\cal C}_a(u)}\oint_{{\cal C}_b(v)}u^{-k}v^{-l}\omega(u,v). \\
a_a(u)[k]&=\frac{\epsilon^{k/2}}{2\pi i\sqrt{k}}\oint_{{\cal C}_a(z_a)}z_a^{-k}\omega^{(a)}(u,z_a),\label{eqn:a}
\end{align}
where $\bar{a}$ means the complement of $a$.

Besides, $X_{ab}$ is given in terms of 
\begin{align}
A_a[k,l]=\frac{\epsilon^{k/2}}{2\pi i\sqrt{k}}\oint_{{\cal C}_a(u)}u^{-k}a_a(u)[l] .\label{eqn:A}
\end{align}
\begin{align}
&X_{aa}=A_a(I-A_{\bar{a}}A_a)^{-1} \\
&X_{a\bar{a}}=I-(I-A_a A_{\bar{a}})^{-1} .
\end{align}

Now we calculate $\epsilon$-expansion for a $g=2$ Riemann surface (double torus) with the pinching parameter. $S_1$ and $S_2$ are tori and their Teichm\"{u}ller parameters are $\tau_1$ and $\tau_2$, respectively.

First of all, we remind the definitions of the Weierstrass's zeta function $\zeta(z)$ and elliptic function ${\cal P}(z)$\cite{FK}, for a torus with a Teichm\"{u}ller parameter $\tau$; 
\begin{align}
\zeta(z)=\frac1{z}+\sum_{m,n\in {\mathbb Z}, m^2+n^2\neq 0}\left[ \frac1{z-m-n\tau}+\frac1{m+n\tau}+\frac{z}{(m+n\tau)^2}\right],
\end{align}
\begin{align}
{\cal P}(z)=\frac1{z^2}+\sum_{m,n\in {\mathbb Z}, m^2+n^2\neq 0}\left[ \frac1{(z-m-n\tau)^2}-\frac1{(m+n\tau)^2}\right].
\label{eqn:wp}
\end{align}
In (\ref{eqn:wp}) the last term of summation $\sum (m+n\tau)^{-2}=E_2(\tau)$ is convergent and known as one of the Eisenstein series $E_{2k}=\sum \frac1{(m+n\tau)^{2k}}$, such that
the zeta function will be expanded as $\zeta(z)=\frac1{z}-\sum_{k=1}^{\infty}E_{2k+2}z^{2k+1}$. 
From ${\cal P}(z)=-\frac{d}{dz}\zeta(z)$, also ${\cal P}(z)$ can be expanded as ${\cal P}(z)=\frac1{z^2}+\sum_{k=1}^{\infty}(2k+1) E_{2k+2}z^{2k}$.

%By definition, it is obvious that ${\cal P}(z)$ is periodic function as
%\begin{align}
%{\cal P}(z)={\cal P}(z+m+n\tau), \ \ m,n\in \mathbb Z.
%\end{align}
%Integrating it, one may realize that $\zeta(z)$ is with quasi-periodic properties
%\begin{align}
%\zeta(z)=\zeta(z+m+n\tau)-(\eta_1 m+\eta_{\tau} n)
% \ \ m,n\in \mathbb Z.
%\end{align}
%With the aid of Legendre's equation $\eta_1 \tau-\eta_{\tau}=2\pi i$,
%we conclude that $\eta_1=E_2$ and $\eta_\tau=\tau E_2-2\pi i$.
Defining $P_1(\tau,z)=\zeta(z)-z E_2(\tau)$, the quasi-periodicity is simply expressed as $P_1(\tau,z)=P_1(\tau,z+m+\tau n)+2\pi i n, \ m,n\in\mathbb{Z}$.

%From (\ref{eqn:ndsk1}), 
${\cal P}(z)$ will provide a unique bilinear two-form (which is differential of second kind) for a torus parametrized in the usual way but not normalized.
We see
\begin{align}
\omega^{(a)}(u,v)=({\cal P}(\tau_a,u-v)+E_2(\tau_a)) dudv =:P_2(\tau_a,u-v)dudv \ \ \ (a=1,2),
\label{eqn:b2f}
\end{align}
is appropriately normalized by the quasi-periodicity of $P_1(z)$, (c.f. $P_2=-\frac{d}{dz}P_1$), where $\tau_a$ is the Teichm\"{u}ller parameter of each torus ($\sim S_a$).

Now we are back to genus two Riemann surface.
From (\ref{eqn:a}) and (\ref{eqn:A}), one may have contributions of several orders of $\epsilon$, for $k,l=1,...,4$
\begin{align}
a_a(u)[k]&
=\left(\epsilon^{1/2}P_2(\tau_a,u)du,-\frac{\epsilon P_2'(\tau_a,d)}{\sqrt{2}},\frac{\epsilon^{3/2} P_2''(\tau_a,d)}{2\sqrt{3}},-\frac{\epsilon^2 P_2'''(\tau_a,d)}{12} \right)\\
A_a[k,l]&
=\left(
\begin{array}{cccc}
\epsilon E_2(\tau_a). & 0 &\sqrt{3} \epsilon^2 E_4(\tau_a) & 0 \\
0 & -3\epsilon^2 E_4(\tau_a) & 0 & -5\sqrt{2}\epsilon^3 E_6(\tau_a) \\ \sqrt{3} \epsilon^2 E_4(\tau_a)  & 0 & 10 \epsilon^3 E_6(\tau_a) & 0 \\ 
0 &  -5\sqrt{2}\epsilon^3 E_6(\tau_a)  & 0 & -35 \epsilon^4 E_8(\tau_a)
\end{array}
\right)
\end{align}

%From (\ref{eqn:omega2})
Then a foundation of our analysis is given;
\begin{align}
\omega(u,v)=\begin{cases}&P_2(\tau_a,u-v)du dv+\epsilon^2 E_2(\tau_{\bar{a}})P_2(\tau_a,u)P_2(\tau_a,v) du dv\\
&+\frac{\epsilon^4}2\left(-3  E_4(\tau_{\bar{a}})P_2'(\tau_a,u)P_2'(\tau_a,v)+E_4(\tau_{\bar{a}}) P_2''(\tau_a,u)P_2'(\tau_a,v)+E_4(\tau_{\bar{a}})P_2'(\tau_a,u)P_2''(\tau_a,v)\right. \\
& \left.+2 E_2(\tau_a) E_2(\tau_{\bar{a}})^2\right)du dv+O(\epsilon^{11/2}), \ \ (u\in S_a, \ v\in S_a)\\
& \\
&-\epsilon P_2(\tau_a,u) P_2(\tau_{\bar{a}},v)du dv -\frac{\epsilon^2}2P_2'(\tau_a,u) P_2'(\tau_{\bar{a}},v) du dv \\
& +\epsilon^3 \left(-\frac1{12}P_2''(\tau_a,u) P_2''(\tau_{\bar{a}},v)-E_2(\tau_a) E_2(\tau_{\bar{a}})P_2(\tau_a,u) P_2(\tau_{\bar{a}},v)\right)du dv \\
&-\frac{\epsilon^4}{144}P_2'''(\tau_a,u) P_2'''(\tau_{\bar{a}},v)du dv+\epsilon^5(-\frac12E_4(\tau_a)E_2(\tau_{\bar{a}})P_2(\tau_a,u)P_2''(\tau_{\bar{a}},v)
\\&-\frac12E_4(\tau_{\bar{a}})E_2(\tau_a)P_2(\tau_{\bar{a}},v)P_2''(\tau_a,u)-
E_2(\tau_a)^2E_2(\tau_{\bar{a}})^2P_2(\tau_a,u)P_2(\tau_{\bar{a}},v)\\
&-3E_4(\tau_a)E_4(\tau_{\bar{a}})P_2(\tau_a,u)P_2(\tau_{\bar{a}},v))dudv+O(\epsilon^{11/2}), \ \  (u\in S_a, \ v\in S_{\bar{a}})
\end{cases}.
\end{align}

%Then for $u\in S_1, v\in S_2$
From the Riemann bi-linear relations, we see the holomorphic one-form for the double torus with pinching parameter $\epsilon$,
\begin{align}
\nu_1(u\in S_1)&= \frac1{2\pi i}\int_{B_1(v)}\omega(u\in S_1,v)\\
&=du+ \epsilon^2 E_2(\tau_2)P_2(\tau_1,u)du+ \frac{\epsilon^4}2 \left( E_4(\tau_2) P_2''(\tau_1,u)+2 E_2(\tau_1)E_2(\tau_2)^2P_2(\tau_1,u)\right)du\\
\nu_1(u\in S_2)& =\frac1{2\pi i}\int_{B_1(v)}\omega(u\in S_2,v) \\
&=-\epsilon P_2(\tau_2,u) du -\epsilon^3 E_2(\tau_1)E_2(\tau_2)P_2(\tau_2,u)du-3\epsilon^5E_4(\tau_2)E_4(\tau_1)P_2(\tau_2,u)du.
\end{align}
As expected, they coincides to that of torus with vanishing $\epsilon$.
Then we can identify its complex structure from them. 
%For example\cite{IT}, the holomorphic one-form determines a period matrix by
%\begin{align}
%&\int_{A_j}\nu_i=\delta_{ij} \\
%\Omega_{ij}&=\int_{B_j}\nu_i\\
%&=\left(
%\begin{array}{cc}
%\int_{B_1}\nu_1=\tau_1+2\pi i\epsilon^2 E_2(\tau_2)+2\pi i \epsilon^4E_2(\tau_1)E_2(\tau_2)^2& \int_{B_2}\nu_1=-2\pi i\epsilon-2\pi i\epsilon^3 E_2(\tau_1)E_2(\tau_2) \\
%\int_{B_1}\nu_2=-2\pi i\epsilon-2\pi i\epsilon^3 E_2(\tau_1)E_2(\tau_2)  & \int_{B_2}\nu_1=\tau_2+2\pi i\epsilon^2 E_2(\tau_1)+2\pi i \epsilon^4E_2(\tau_2)E_2(\tau_1)^2 \\
%\end{array}
%\right) .
%\end{align}
By the way, in our analysis $\tau_1,\tau_2$ are set to $i$ as a standard one for a background torus providing $E_2(i)=0$.
The holomorphic quadratic differentials can be given by $\varphi_{ij}=\nu_i\nu_j$ as,
\begin{align}
\varphi_{11}&=
\begin{cases}
\left(1+\epsilon^4 E_4(i)P_2''(i,z)\right)dz^2 +O(\epsilon^6) & z\in S_1 \\
 \epsilon^2 P_2(i,z)^2 dz^2 +6\epsilon^6 E_4(i)^2 P_2(i,z)^2 dz^2 +O(\epsilon^6)\in S_2 \\
\end{cases}  \\
\varphi_{12}&=
\begin{cases}
-\epsilon P_2(i,z)dz^2 +\epsilon^5\left(-3E_4(i)^2P_2(i,z)^2-\frac12E_4(i)P_2''(i,z)P_2(i,z) \right)dz^2+O(\epsilon^6)& z\in S_1 \\
-\epsilon P_2(i,z)dz^2 +\epsilon^5\left(-3E_4(i)^2P_2(i,z)^2-\frac12E_4(i)P_2''(i,z)P_2(i,z) \right)dz^2+O(\epsilon^6) & z\in S_2 \\
\end{cases} 
\end{align}
and their permutation$(1\leftrightarrow 2)$.

%---------   References   ---------%

\end{document}